\documentclass[draft, onecolumn]{IEEEtran}
\usepackage{amsfonts, amsthm}
\usepackage[cmex10]{amsmath}
\usepackage{graphicx}
\usepackage{multicol}
\usepackage{multirow}
\usepackage{amssymb}
\usepackage{lscape}
\usepackage[english]{babel}
\usepackage{cite}
\usepackage{bbm}
\usepackage{color}
\usepackage{soul}
\usepackage{tikz, adjustbox}
\usetikzlibrary{automata,arrows,positioning,calc}

\newcommand{\Exp}{\mathsf E}
\newcommand{\expect}[1]{{\Exp}\left[#1\right]}
\newcommand{\expcnd}[2]{{\Exp}\left[\left. #1 \,\right|\, #2\right]}

\newcommand{\X}{{\bf X}}
\newcommand{\x}{{\bf x}}
\newcommand{\MC}[1]{$\{\X_t^{(#1)}\}$}

\newtheorem{theorem}{Theorem}[section]
\newtheorem{lemma}{Lemma}[section]
\newtheorem{corollary}{Corollary}[section]

\newcommand{\be}{\begin{equation}}
\newcommand{\ee}{\end{equation}}

\title{Stability and Instability Conditions for Slotted Aloha with Exponential Backoff}

\author{L.~Barletta,~\IEEEmembership{Member,~IEEE}, and F.~Borgonovo,~\IEEEmembership{Member,~IEEE}.
	\thanks{Manuscript received Month DD, 2017; revised Month DD, 2017.}% <-this % stops a space
	\thanks{L.~Barletta and F.~Borgonovo are with the Dipartimento di Elettronica,
		Informazione e Bioingegneria, Politecnico di Milano, I-20133 Milano, Italy (E-mail: \{luca.barletta, flaminio.borgonovo\}@polimi.it).} %
	\thanks{Part of this work was submitted to the 2017 IEEE International Symposium on Information Theory.}
}%

\begin{document}

\maketitle

\maketitle\begin{abstract}
This paper provides  stability and instability conditions for slotted Aloha under the exponential backoff (EB) model with   geometric law $i\mapsto b^{-i-i_0}$, when transmission buffers are in saturation, \emph{i.e.}, always full. In particular, we prove that for any number of users and for $b>1$ the system is: (i) ergodic for $i_0 >1$, (ii) null recurrent for $0<i_0\le 1$, and (iii) transient for $i_0=0$. Furthermore, when referring to a system with queues and Poisson arrivals, the system is shown to be stable whenever EB in saturation is stable with throughput $\lambda_0$ and the system input rate is upper-bounded as $\lambda<\lambda_0$.
\end{abstract}


%---------------------------------------------------
\section{Introduction}
 	\label{sec:intro}
 	
The Aloha protocol, since its appearance in 1970 \cite{aloha1}, has been perhaps the  most studied subject in the multiple-access area, since its introduction  	has revolutionized the multiple-access world. Its applications cover  	important fields such as satellite, cellular and local-area communications, recently being applied to radio frequency identification. However, many  	questions regarding its stability, under different circumstances and channel assumptions, remain unanswered, especially when dealing with an  	unlimited number of users and  no channel feedback is available, as is the case  	here considered.
 	
Aloha and its slotted version S-Aloha \cite{roberts}, in their basic versions with constant retransmission probability $\beta$, have soon be proved unstable  	\cite{lam},\cite{gelembe}.  Assuming that some additional information is available  	from the observation of the channel, an estimate $\widehat{N}$ of the number of users $N$ can be attempted and the throughput optimized by setting $\beta=1/\widehat{N}$. For example, in some broadcast channels, transmissions can be monitored  and  the outcome of the slot, \emph{i.e.},  \emph{empty}, \emph{success}, or \emph{collided} is made available. In  \cite{lam, hajek, rivest, clare} procedures and estimates have been suggested that are able to provide  the theoretical throughput of $e^{-1}$.
 	
When  no channel feedback is available, the retransmission probability must be set with other means. Since a constant retransmission probability can not stabilize the protocol, further studies must consider a retransmission probability that changes according to the user's own history.  The only mechanism of this type so far considered, called \textit{backoff},  reduces the retransmission probability $\beta(i)$ as the number of collisions $i$ suffered by the packet increases, on the ground that the number of suffered collisions provides a measure of the channel congestion degree.

The mechanism most often referred to is the  exponential backoff (EB), which decreases the user's transmitting probability according to the negative exponential law
 \be \beta(i)=b^{-i-i_0}, \label{eq: MBEB}\ee
where $i\ge 0$ counts the  number of consecutive collisions experienced in transmitting a packet, the exponential base is $b>1$, and $i_0$ is the transmission probability offset of the first attempt.  EB has undergone many analysis efforts in order to assess its capabilities, especially in view of the fact that, with some variations, it has been adopted in IEEE 802.3 and IEEE 802.11 standards, in the binary EB (BEB) variation, \emph{i.e.}, with $b=2$.  Many efforts have been devoted to investigate issues such as its stability and throughput. Unfortunately, the analysis of such protocol is quite complex and the  results attained are partial and somehow contradictory and confusing, owing to the many differences in the assumptions underlying the  analyzed models and  stability definitions.

\subsection*{State of the Art}
%\noindent \emph{\bf State of the Art:} 
The BEB for S-Aloha under the infinite population model has been proved unstable by Aldous in \cite{aldous}. There, the author considers the infinite population model where users arrive, transmit their packet according to law (\ref{eq: MBEB}) with $i_0=0$, and after success leave. The author proves that such model is unstable under any positive packet-arrival rate $\lambda$, so that its throughput is zero.

Subsequent papers have tried to analyze a model composed of $N$ users with BEB, $i_0=0$, and Markovian arrivals that await their transmission turn in a local queue.  Here the variable involved are the backoff indexes at each station, $X_1,X_2,\ldots, X_N$, and the content of each queue, $Q_1,Q_2,\ldots, Q_N$,  where $N$ can be unbounded. Due to the complexity of the model exact analyses have been never produced, and only bounds on the throughput have been attained. In \cite{goodman}, Goodman \emph{et al.} prove that an arrival frequency $\lambda^*(N)>0$ does exist such that the system is stable if  $\lambda(N) < \lambda^*(N)$, where $\lambda^*(N) \ge 1/N^{\alpha \log N}$ for some constant $\alpha$. In \cite{ammal}, Al-Ammal \emph{et al.} improve the bound in \cite{goodman} proving that BEB is stable for arrival rates smaller than $1/\alpha N^{1-\eta}$, where $\eta < 0.25$. Finally, in \cite{hastad}, H\aa stad \emph{et al.} show, using the same analytical model as in \cite{goodman}, that BEB is unstable whenever $\lambda_i > \lambda/N$ for $1 \le i \le N$,  and $\lambda >0.567 + 1/(4N-2)$, where $\lambda$ is the system arrival rate and $\lambda_i$ is the arrival rate at node $i$. 
 	
Due to the complexity of an exact analysis, further attempts have introduced simplified models and approximations. Among these, the saturation model has been first introduced in \cite{bianchi}. This model tries to analyze stability and throughput issues by assuming that queues are always full, such that, once a successful transmission has occurred at a station, immediately a new one is available for transmission. This model is somewhat simpler and pessimistic with respect to the one with queues, and has been adopted in the hope that it presented a stable behavior and positive throughput, thus guaranteeing the stable behavior and the throughput of the more realistic one.
 	
With the saturation model, an approximate analysis is made possible by the \textit{decoupling assumption} \cite{bianchi},  \cite{byung}. This assumption has a twofold implication, \emph{i.e.}, the \textit{stationary} behavior of the model, and the \textit{independence} in the behavior of the different transmitters. These implications lead to a mean value analysis (MVA) and a fixed point equation that yields the basic performance figures of the protocol. This analysis provides acceptable results when using large $i_0$ and a finite number of backoff stages, but largely underestimates the throughput with low $i_0$ and an infinite number of backoff stages, the only case able to deal with unlimited $N$. In \cite{infocom-2016}, we have introduced a new model, still with queues in saturation, that very closely approximates the behavior of the real system. In all cases, however, no formal proof of the stability conditions has been given.

\subsection*{Contributions}
%\noindent
%\emph{\bf Contributions:} 
In this paper we investigate the stability of EB,  as defined by~\eqref{eq: MBEB}, with an unlimited number of backoff stages. First we consider the saturation model, and derive the conditions under which the Markov chain (MC) modeling the system is positive recurrent, null recurrent, and transient. In particular, we  prove that the EB is ergodic only for  $b>1$ and $i_0 > 1$, null recurrent for $0<i_0\le 1$, and transient for $i_0=0$. When transient, all indexes but one increase, leading to the phenomenon known as channel capture where, in the end, only one user successfully transmits with probability $1$. Furthermore, we prove that some backoff indexes, in addition to the lowest one, reach a stationary behavior even in the null-recurrent case, showing that a group of users capture the channel for an infinite time in the average, while the others are locked out. We finally prove that for $i_0 \le 1/(N-1)$ all indexes but the lowest one diverge, and the throughput is one. 

Finally, we show that, when queues are considered, under a Poisson arrival process at rate $\lambda$, the joint occupancy process is ergodic for $b>1$,  $i_0 > 1$, and an arrival rate $\lambda <\lambda_0$, where $\lambda_0$ is the throughput of EB in saturation conditions.

We must note that the fact that backoff can reach $100\%$ throughput, albeit with capture, is not a completely new result since in \cite{hastad} this property was proven for a polynomial backoff law of the type
\be \beta(i)=(i+1)^{-\alpha} \label{eq: PBEB}\ee
with $\alpha>1$, in a system with queues, Bernoulli arrivals and any number of users. In this paper we give the instability region where a similar result holds for EB with saturated queues.

\subsection*{Organization}
%\noindent
%\emph{\bf Organization:} 
The paper is structured as follows. In Sec.~\ref{Sec:model} we introduce the MC model that describes the EB mechanism, and explain the notation. In Sec.~\ref{Sec:preliminaries} we state and discuss some preliminary results, while in Sec.~\ref{Sec:stability} we expose the main results about stability.  Finally, in Sec.~\ref{Sec:genstability} we provide stability results for the system with queues. Conclusions are given in Sec.~\ref{Sec:conclusions}.

%------------------------------------------------------------------------------------------------------
\section{Backoff Model}\label{Sec:model}
\subsection*{Notation}
%\noindent
%\emph{\bf Notation:} 
Uppercase and lowercase letters, \emph{e.g.} $X$ and $x$, respectively denote random variables (RV)s and their realizations. A similar rule holds for vector-valued variables, \emph{e.g.} $\X$ and $\x$. The probability of the event $\{X=x\}$ is written as $\Pr(X=x)$. The operator $\Exp$ is used for statistical expectations.

Vector-valued random processes are denoted by $\{\X_t\}$ for short. The $i$-th entry of a vector random process at time instant $t$ is denoted with $X_i(t)$.

%The symbol $\ind(\cdot)$ is an indicator function that gives 1 only if the predicate in brackets is true.

\subsection*{Exponential Backoff Model}
%\smallskip
%\noindent
%\emph{\bf Backoff Model:} 
We consider a system with $N$ users whose transmission queues are always full, meaning that after a successful transmission of a packet, another packet is immediately available in the transmission buffer. The state of a user is determined by its backoff index $x$, which is increased at each collision and reset to zero upon a successful transmission. Clearly, with $N$ users, the state of the system at time slot $t$ is the vector $(x_1(t), x_2(t), \ldots, x_N(t))$, where $x_i(t)\le x_j(t)$ for $i\le j$. We denote with $u_i(t)$ a user with back-off index $x_i$ at time $t$.

Transmission of user $u_i(t)$ occurs with probability $b^{-x_i(t)-i_0}$,  and the $i$-th back-off state evolves as a random process $\{X_i(t)\}$, while the system with $N$ users as a vector process $\{\X_t^{(N)}\}=\{(X_1(t), X_2(t), \ldots X_N(t))\}$, or $\{\X_t \}$.

%Note that the back-off law~\eqref{eq: MBEB} is not exactly the window probability law used in standards, where the transmission slot is picked at random in a window $(0,W)$ where $W$ is uniformly distributed in $[1,b^{x_i(t)+i_0}]$. However, as shown in \cite{infocom-2016}, many of the results described  in the literature, such as maximum throughput and average delays, are the same for both the cited back-off mechanisms. In any cases, the memoryless property of~\eqref{eq: MBEB} assures that $\{\X(t)\}_{t\ge 0}$ is an MC, thus amenable to statistical analysis.

%---------------------------------------------------
\section{Preliminary Results}\label{Sec:preliminaries}

In this section we prove some lemmas that are crucial to the proofs of the main theorems of Sec. \ref{Sec:stability}.

% 	The stability of $\{\X_t  \}$  is dictated by the statistics of $\X_t$  after a great number of consecutive collisions. Under this regime, the next lemmas provide some useful bounds. To this purpose,

In the following we consider the set of time instants  ${\bf k}=(k_1,k_2,\ldots, k_j,\ldots)$,  where at least one of the users $\{u_i(0)\}_{i=2}^N$ transmits. The initial state of the MC is ${\bf x}=(0,m,\ldots)$ for a suitable large $m$.

Denote
\begin{align}
\Delta_N(k_j) &= X_N(k_j)-X_2(k_j), \label{eq:boundN2} \end{align}
\vspace{-0.5cm}
\begin{align}
a_0 &= \frac{1}{N-1}\sum_{r=2}^N x_r(0) , \quad
\Delta_C(k_j)=\frac{1}{N-1}\sum_{l=1}^j C(k_l),
\end{align}
where $C(k_l)$ represents number of extra-collisions, \emph{i.e.}, the number of users among  $\{u_r\}_{r=2}^N$ that collides in excess of the one that surely occurs  at time $k_l$. We have the following lemmas.
\begin{lemma}\label{Lemma:bound1}
For any $b>1$ and $i_0\ge 0$,  the following bounds hold for $N\ge 2$:
\begin{align}
-\Delta_N(k_j) \le X_2(k_j)-a_0 - \frac{j}{N-1} \le \Delta_C(k_j). \label{eq: bounduno}
\end{align}
\end{lemma}
\begin{IEEEproof}
Since by hypothesis in all time instants ${\bf k}$ the transmitting users collide, possibly with $u_1$, the components of $\X$ always increase, and we can write
\be
\sum_{r=2}^N X_r(k_j)= \sum_{r=2}^N x_r(0) +j+ \sum_{l=1}^j C(k_l), \ee
from which bounds \eqref{eq: bounduno} immediately descend.
\end{IEEEproof}

\begin{lemma}\label{Lemma:bound2}  For any $b>1$, $i_0\ge 0$, and for $N\ge 2$, we have
\be  \Pr(\Delta_N(k_j)= \delta)\le{\cal O}(b^{-\delta^2}), \qquad  \mbox{as \ } \delta\rightarrow \infty,
\label{eq:distr11}\ee
and
\be  \expect{b^{\Delta_N(k_j)}} \le D, \qquad \forall k_j
\label{eq:distr12}\ee
where $D$ is a finite constant.
\end{lemma}
\begin{IEEEproof}
Let refer to process  $\{\X_{K_{j}}\}$ in the form  $(\Delta_N,\widetilde{\X} )=(\Delta_N,X_2,X_3,\ldots,X_{N-1})$ and denote by $N_2$ and $N_N$ the number of users whose index is equal to $X_2$ and $X_N$, respectively. $\Delta_N$ increases if at least one of the $N_N$ users transmits and  at least one of the $N_2$ users does not transmit.
Therefore, the corresponding probability is
\begin{align}
&\Pr(\Delta_N(k_{j+1})=\delta+1| \Delta_N(k_j)=\delta, N_2(k_j)=n_2, \widetilde{\X}_{k_j}=\x)\nonumber\\
&= \frac{(1-(1- b^{-x_N-i_0})^{n_N}) (1-(b^{-x_2-i_0})^{n_2}) }{1-\prod_{i=2}^{N}(1-b^{-x_i-i_0})} \nonumber\\
&{\le} \frac{n_N\: b^{-x_N-i_0} }{1-(1-b^{-x_2-i_0})}\le (N-1)\: b^{-\delta},  \qquad \forall \x\in {\cal A}. \label{eq:tran2a}
\end{align}
where ${\cal A}(\x)$ is the set of states ${\bf x}$ that are compatible with $n_2$.

$\Delta_N$ decreases if all the $N_2$ users transmit and none of the  $N_N$ users transmits. We can write the following lower bound:
\begin{align}
&\Pr(\Delta_N(k_{j+1})=\delta-1| \Delta_N(k_j)=\delta, N_2(k_j)=n_2, \widetilde{\X}_{k_j}=\x) \nonumber \\
&= \frac{(b^{-x_2-i_0})^{n_2}(1-b^{-x_N-i_0})^{n_N}}{1-\prod_{i=2}^{N}(1-b^{-x_i-i_0})}\nonumber\\
&\ge \left\{
\begin{array}{lc}
\frac{b^{-x_2-i_0}(1-b^{-x_N-i_0})^{n_N}}{1-(1-b^{-x_2-i_0})^{N-1}} & n_2=1, \qquad \forall \x\in {\cal B}, \\
0 & n_2>1\qquad \forall \x\in {\cal C},
\end{array}
\right. \nonumber\\
&\ge \left\{
\begin{array}{lc}
\frac{(1-b^{-x_2-i_0})^{N-1} }{N-1} & n_2=1, \qquad \forall \x\in {\cal B}, \\
0 & n_2>1\qquad \forall \x\in {\cal C},
\end{array}
\right.
\label{eq:tran3}
\end{align}
where ${\cal B}$ and ${\cal C}$ are respectively the set of states $\x$ that are compatible with $n_2=1$ and $n_2>1$ respectively.

We now build process  $(\Delta'_N,\X')=(\Delta'_N,X'_2,X'_3,\ldots,X'_{N-1})$ derived from $(\Delta_N,\widetilde{\X})=(\Delta_N,X_2,X_3,\ldots,X_{N-1})$, where the transition probabilities
\[ \Pr(\Delta_N(k_{j+1})=\delta\pm 1, \widetilde{\X}_{k_{j+1}}={\bf y}| \Delta_N(k_j)=\delta, \widetilde{\X}_{k_j}=\x)\]
are replaced by the transition probabilities
\[ \Pr(\Delta'_N(k_{j+1})=\delta \pm 1, \X'_{k_{j+1}}={\bf y}| \Delta'_N(k_j)=\delta, \X'_{k_j}=\x)\]
that can be chosen rather arbitrarily but for the constraints
\begin{align}
&\Pr\Big(\Delta'_N(k_{j+1})=\delta+1, N'_2(k_{j+1})=a \, \Big|\nonumber\\
&\qquad \Delta'_N(k_j)=\delta, N'_2(k_j)=c, \X'_{k_j}=\x \Big)\nonumber\\
&= \left\{
\begin{array}{lc} \min\{(N-1)\: b^{-\delta},1\}, \qquad &  a=1, \ \forall \x,
\\ 0, \qquad & \text{elsewhere}
\end{array}
\right.\label{eq:tran2a}
\end{align}
\begin{align}
&\Pr\Big(\Delta'_N(k_{j+1})=\delta-1, N'_2(k_{j+1})=a \Big| \nonumber\\
& \qquad\Delta'_N(k_j)=\delta, N'_2(k_j)=c, \X'_{k_j}=\x\Big)  \nonumber \\ &=\left\{
\begin{array}{lc}
\frac{(1-b^{-x_2-i_0})^{N-1} }{N-1} & \qquad  c=a=1, \\
0 & \qquad  \text{elsewhere}
\end{array}
\right. \label{eq:tran3a}
\end{align}
\begin{align}
&\Pr\Big(\Delta'_N(k_{j+1})=\delta, N'_2(k_{j+1})=a \Big|\nonumber\\
&\qquad \Delta'_N(k_j)=\delta, N'_2(k_j)=c, \X'_{k_j}=\x \Big)   \nonumber \\ &=\left\{
\begin{array}{lc}
\gamma_{\delta,c} & \qquad   \ a = c-1,\, c>1 \\
0 & \qquad  \mbox{\ elsewhere}
\end{array}
\right. \label{eq:tran4a}
\end{align}
where $\gamma_{\delta,c}$ is such that the summation of transition probabilities over $\delta-1, \delta,$ and $\delta+1$ is one.

If in \eqref{eq:tran2a} and \eqref{eq:tran3a} we take the summation over $a$ we get
\begin{align}
&\Pr(\Delta'_N(k_{j+1})=\delta+1| \Delta'_N(k_j)=\delta, N'_2(k_j)=c, \X'_{k_j}=\x)\nonumber\\
&=  \min\{(N-1)\: b^{-\delta},1\}  \nonumber  \\
& \ge \Pr(\Delta_N(k_{j+1})=\delta+1| \Delta_N(k_j)=\delta, N_2(k_j)=c, \widetilde{\X}_{k_j}=\x),
\label{eq:tran22a}
\end{align}
\begin{align}
&\Pr(\Delta'_N(k_{j+1})=\delta-1| \Delta'_N(k_j)=\delta, N'_2(k_j)=c, \X'_{k_j}=\x)  \nonumber \\
%&= \frac{(1-b^{-x_2-i_0})^{N-1} }{N-1} , \qquad \forall \x, \nonumber \\
&=\left\{
\begin{array}{lc}
\frac{(1-b^{-x_2-i_0})^{N-1} }{N-1} & \qquad  c=1 \\
0 & \qquad  c>1
\end{array}
\right.
\nonumber\\
& \le \Pr(\Delta_N(k_{j+1})=\delta-1| \Delta_N(k_j)=\delta, N_2(k_j)=c, \widetilde{\X}_{k_j}=\x)
\label{eq:tran33a}
\end{align}
for all $\x$.

Inequalities \eqref{eq:tran22a} and \eqref{eq:tran33a}  satisfy the conditions \eqref{eq: delta1} and \eqref{eq: delta2} of Lemma \ref{lem:delta_dominanza} in Appendix~\ref{App:MC}. Condition  \eqref{eq: delta3} is met by forcing $\Delta'_N(k_j)\ge \delta^*$, where $\delta^*$ is  taken so high as to satisfy $(N-1)\: b^{-\delta^*} <1/2$. Therefore, the marginal process $\Delta'_N$ stochastically dominates $\Delta_N$, \emph{i.e.},
\be  \Pr(\Delta_N'(k_j) > \delta)\ge \Pr(\Delta_N(k_j) > \delta), \qquad \delta=0,1,\ldots.
\label{eq:distr111}\ee
Furthermore, $(\Delta'_N(k_j),N_2(k_j))$ is itself a Markov Chain, whose transition diagram is shown in Fig.~\ref{fig1.3}, with
\begin{align} \alpha_{\delta}&=(N-1)\: b^{-\delta}  & \qquad \delta \ge \delta^*   \nonumber \\
\beta_\delta & =\beta=\frac{(1-b^{-x_2-i_0})^{N-1} }{N-1}  & \qquad \delta \ge \delta^*+1,
\end{align}
and is such that states with $N_2>1$ are transient. Then, the asymptotic distribution can be evaluated only referring to positive-recurrent states, \emph{i.e.}, to Markov Chain  $\{\Delta'_N(k_j)\}$, where we implicitly assume $N_2(k_j)=1$ for all $k_j$'s.
$\{\Delta'_N(k_j)\}$ is the well known discrete-time Birth \& Death process, whose  birth and death rate are $ \alpha_{\delta}$ and $\beta$.
Its asymptotic distribution is given by
\begin{align} \pi_{\delta^*+\delta} &=  \pi_{\delta^*}\:   \frac{\alpha_{\delta^*}\alpha_{\delta^*+1} \ldots \alpha_{\delta^*+\delta-1}}{\beta^\delta}  \nonumber \\
&= \pi_{\delta^*}\:  \left( \prod_{k=0}^{\delta-1} b^{-k-\delta^*} \right) \left(\frac{(N-1)^2}{1-b^{-x_2-i_0}}\right)^\delta \nonumber \\
&= \pi_{\delta^*}\:   b^{-\delta(\delta-1)/2-\delta\cdot\delta^*}  \left(\frac{(N-1)^2}{1-b^{-x-i_0}}\right)^\delta
\end{align}
for $\delta>0$, which is ${\cal O}(b^{-\delta^2})$ for $\delta\rightarrow \infty$ thus proving \eqref{eq:distr11}.

%\begin{figure}[t]
%%\vspace{0.5cm} \centering
% \epsfxsize=11cm
%  \centering  {\epsffile{figure/diagram-2b.eps}}
%\caption{\em  }
% \label{fig1.3} %\vspace{-0.5cm}
%\end{figure}

% 		\begin{figure}
% 			\centering
% 			\includegraphics[width=.75\columnwidth]{figure/diagram-2b}
% 			\caption{\em}\label{fig1.3}
% 		\end{figure}
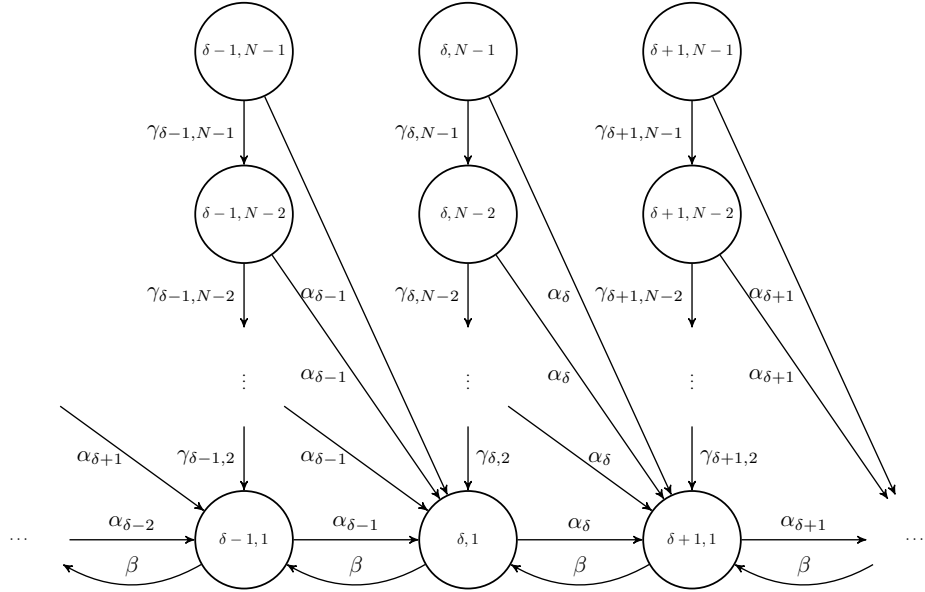
\begin {figure*}%[!hbtp]
\centering
\begin{adjustbox}{width=0.8\textwidth}
\begin{tikzpicture}[->, >=stealth', auto, semithick, node distance=5cm]
\tikzstyle{every state}=[fill=white,draw=black,thick,text=black,scale=0.7, minimum size=6.2em]
\node[state]    (A1)                     {$\delta-1,N-1$};
\node[state]    (A2)[below=1cm of A1]        {$\delta-1,N-2$};
\node[state, draw=none]    (A3)[below=1cm of A2]        {$\vdots$};
\node[state]    (A4)[below=1cm of A3]        {$\delta-1,1$};
\node[state]    (B1)[right of=A1]        {$\delta,N-1$};
\node[state]    (B2)[below=1cm of B1]        {$\delta,N-2$};
\node[state, draw=none]    (B3)[below=1cm of B2]        {$\vdots$};
\node[state]    (B4)[below=1cm of B3]        {$\delta,1$};
\node[state]    (C1)[right of=B1]        {$\delta+1,N-1$};
\node[state]    (C2)[below=1cm of C1]        {$\delta+1,N-2$};
\node[state, draw=none]    (C3)[below=1cm of C2]        {$\vdots$};
\node[state]    (C4)[below=1cm of C3]        {$\delta+1,1$};
\node[state, draw=none] (3) [left of=A3] {};
\node[state, draw=none] (4) [left of=A4] {$\cdots$};
\node[state, draw=none] (D4) [right of=C4] {$\cdots$};
\path
(A1) edge[left]     node{$\alpha_{\delta-1}$}         (B4)
(A2) edge[left]     node{$\alpha_{\delta-1}$}         (B4)
(A3) edge[left]     node{$\alpha_{\delta-1}$}         (B4)
(A4) edge[above]     node{$\alpha_{\delta-1}$}         (B4)
(B1) edge[left]     node{$\alpha_{\delta}$}         (C4)
(B2) edge[left]     node{$\alpha_{\delta}$}         (C4)
(B3) edge[right]     node{$\alpha_{\delta}$}         (C4)
(B4) edge[above]     node{$\alpha_{\delta}$}         (C4)
(A1) edge[left]     node{$\gamma_{\delta-1,N-1}$}         (A2)
(A2) edge[left]     node{$\gamma_{\delta-1,N-2}$}         (A3)
(A3) edge[left]     node{$\gamma_{\delta-1,2}$}         (A4)
(B1) edge[left]     node{$\gamma_{\delta,N-1}$}         (B2)
(B2) edge[left]     node{$\gamma_{\delta,N-2}$}         (B3)
(B3) edge[right]     node{$\gamma_{\delta,2}$}         (B4)
(C1) edge[left]     node{$\gamma_{\delta+1,N-1}$}         (C2)
(C2) edge[left]     node{$\gamma_{\delta+1,N-2}$}         (C3)
(C3) edge[right]     node{$\gamma_{\delta+1,2}$}         (C4)
(B4) edge[bend left, above]     node{$\beta$}         (A4)
(C4) edge[bend left, above]     node{$\beta$}         (B4)
(A4) edge[bend left, above] 	node{$\beta$}		(4)
(D4) edge[bend left, above] 	node{$\beta$}		(C4)
(C4) edge[above] 	node{$\alpha_{\delta+1}$}		(D4)
(4) edge[above] 	node{$\alpha_{\delta-2}$}		(A4)
(C1) edge[left] 	node{$\alpha_{\delta+1}$}		(D4)
(C2) edge[left] 	node{$\alpha_{\delta+1}$}		(D4)
(3) edge[left] 	node{$\alpha_{\delta+1}$}		(A4);
\end{tikzpicture}
\end{adjustbox}
\caption{Auxiliary Markov chain $(\Delta_N',N_2)$.} \label{fig1.3}
\end{figure*}

 		For large $\delta_0$ we can write
 		\begin{align}
 		\expect{b^{\Delta'_N(k_j)}} &= A + \sum_{\delta=\delta_0+1}^\infty \Pr(\Delta_N'(k_j)=\delta)\: b^\delta \nonumber\\
 		&\le A + \beta \sum_{\delta=\delta_0+1}^\infty b^{-\delta^2+\delta} =D  \label{eq:bound_bDelta}
 		\end{align}
 		where $A$, $\beta$, and $D$ are finite constant. Here, the inequality follows from stochastic dominance, since $b^x$ is an increasing function of $x$ \cite{quirk1962}.
 		
 	\end{IEEEproof}

 	\begin{lemma}\label{Lemma:bound3}
 		For any $b>1$ and $i_0\ge 0$, and for $N\ge 2$, we have
 		\be  \expect{b^{\Delta_C(k_j)}} < \infty.
 		\label{eq:distr13}\ee
 	\end{lemma}
\begin{IEEEproof}
	Let consider a realization path $\{\delta_N(k_j),\x_{k_j}\}$ of the process $\{\Delta_N(k_j),\widetilde{\X}_{k_j} \}$.  The number of extra-collisions in a path is stochastically dominated by RV $(N-1)\Delta'_C(k_j)$,  number of extra-collisions attained assuming $x_i'(k_j)=a_j, \ 2 \le i \le N$, where, by Lemma~\ref{Lemma:bound1},
	\be x_2(k_j)\ge a_j = a_0 + \frac{j}{N-1}-\delta_N(k_j). \label{eq:alpha_j} \ee
	
	%In fact, owing again to the property of statistical dominance \cite{xx}, we have
	%\be \expect{b^{-X_2(k_j)-i_0}} \le \expect{b^{-\alpha_j-i_0}}. \ee
	With this assumption, the number of extra-collisions at time $k_j$, $C'(k_j)$, is a binomial variable between $0$ and $N-2$ with success probability $b^{-a_j-i_0}$. %If we consider a given path $\{\delta_N(k_j)\}$ of Markov Chain $\{\Delta_N(k_j)\}$,
	The RVs $C'(k_j)$ at different $k_j$ are statistically independent and each of them has $z$-transform
	\be {\cal B}_j(z)=( 1+b^{-a_j-i_0}(z-1)  )^{N-2}.  \ee
	Let now consider the total number of extra-collisions
	\be  (N-1) \Delta'_C(k_j)=\sum_{\ell=1}^j C'(k_\ell), \ee
	whose $z$-transform is, by independence of the $C'(k_\ell)$'s,
	\begin{align}{\cal D}(z)& =\prod_{\ell=1}^j {\cal B}_\ell(z).
	%= \prod_{j=1}^\infty (b^{-\alpha_j-i_0} z + 1-b^{-\alpha_j-i_0})^{N-2} \\
	%& = \prod_{\ell=1}^j (b^{-\alpha_\ell-i_0} z + 1-b^{-\alpha_\ell-i_0})^{N-2}.
	\end{align}
	The expectation with respect to $\Delta'_C(k_j)$ conditioned to $\{\Delta_N(k_\ell)\}=\{\delta_N(k_\ell)\}$
	\begin{align} \expcnd{b^{\Delta'_C(k_j)}}{ \{\delta_N(k_\ell)\}}&= \expcnd{(b^{\frac{1}{N-1}})^{(N-1)\Delta'_C(k_j)}}{\{\delta_N(k_\ell)\}} \end{align}
	%\begin{align} \Exp_{\Delta'_C}[{b^{\Delta'_C(k_j)}}]&= \Exp_{\Delta'_C}\left[(b^{1/(N-1)})^{(N-1)\Delta'_C(k_j)}\right]\end{align}
	coincides with the $z$-transform ${\cal D}(z)$ evaluated in $z=b^{1/(N-1)}$. Therefore we have
	\begin{align} &\ln\expcnd{b^{\Delta'_C(k_j)}}{\{\delta_N(k_\ell)\}} \nonumber\\
	&\quad= (N-2)\sum_{\ell=1}^j  \ln\left( 1+b^{-a_\ell-i_0}(b^{\frac{1}{N-1}}-1) \right)\nonumber \\
	&\quad\le (N-2) (b^{\frac{1}{N-1}}-1) \sum_{\ell=1}^j b^{-a_0-\frac{\ell}{N-1}+\delta_N(k_\ell)-i_0}\nonumber\\
	&\quad\le (N-2) (b^{\frac{1}{N-1}}-1) \sum_{\ell=1}^\infty b^{-a_0-\ell (\frac{1}{N-1}-\frac{\delta_N(k_\ell)}{\ell})-i_0}. \label{eq:convDeltaC}
	%\nonumber\\
	%&=  \left(\prod_{\ell=1}^j  \left( 1+b^{-a_j-i_0}(b^{1/(N-1)}-1)  \right) \right)^{N-2}.
	\end{align}
	
	One of the consequences of result~\eqref{eq:distr11} of Lemma~\ref{Lemma:bound2} is that
	\begin{align}
	\Pr\left(\Delta_N(k_\ell)>\epsilon \ell \right) = \beta\: b^{-(\epsilon \ell)^2}
	\end{align}
	for a suitably large value of $\ell$ and for all $\epsilon>0$.
	This implies that the condition of the Borel-Cantelli lemma
	\begin{align}
	\sum_{\ell = 1}^{\infty} \Pr\left(\frac{\Delta_N(k_\ell)}{\ell}>\epsilon \right) < \infty
	\end{align}
	is satisfied for all $\epsilon>0$, which tells that the events \mbox{$\{ \frac{\Delta_N(k_\ell)}{\ell}>\epsilon \}_\ell$} occur finitely many times. At this point we can write
	\begin{align}
	\Pr\left( \lim_{\ell\rightarrow\infty} \frac{\Delta_N(k_\ell)}{\ell} > \epsilon \right) = 0, \qquad \forall \epsilon>0,
	\end{align}
	or
	\begin{align}\label{eq:BorelCantelli}
	\Pr\left( \lim_{\ell\rightarrow\infty} \frac{\Delta_N(k_\ell)}{\ell} = 0 \right) = 1.
	\end{align}
	Using the convergence~\eqref{eq:BorelCantelli} and the continuous mapping theorem, we can claim that the series in~\eqref{eq:convDeltaC} converges almost surely to a geometric series, which converges to a finite value. Therefore we can write
	\be  \expect{b^{\Delta'_C(k_j)}} < \infty. \ee  The thesis follows by stochastic dominance.
\end{IEEEproof}

 	% Lemma -----------------------------------------
 	
 	\begin{lemma}\label{Lemma1}
 		If in all time instants ${\bf k}$ the transmitting users collide, then the probability that $X_1=0$ at all times in  ${\bf k}$ is non-zero. That is, denoting with ${\cal C}(\x)$ the channel collision event when the indexes are $\x$, we have
 		\begin{equation}\label{eq:thesis}
 		\expect{\prod_{j=1}^{\infty}\Pr\left(X_1(K_j)=0| \{X_1(K_v)=0,  {\cal C}(\X_{K_v})\}_{v=1}^{j-1} \right) } >0 \nonumber
 		\end{equation}
 		for $b>1$ and some $x_2(k_1)=m>0$.
 	\end{lemma}
 	\begin{IEEEproof}
 		We have
 		\begin{align}
 		c(k_{j},k_{j+1}) & \triangleq  \Pr(X_1(k_{j+1})=0|\{X_1(k_{l})=0, {\cal C}(\X_{k_{l}})\}_{l<j+1}) \nonumber \\
 		&=\Pr(X_1(k_{j+1})=0|X_1(k_{j})=0, {\cal C}(\X_{k_{j}})) \nonumber \\
 		& \ge  \Pr(X_1(k_{j+1})=0|X_1(k_{j}+1)=1)  \nonumber
 		\\ &= 1-(1-b^{-1-i_0})^{k_{j+1}-k_{j}-1},   \label{eq:alpha}
 		\end{align}
 		where the inequality comes from the fact that  $X_1(k_j+1)=1$ only if $u_1$ transmitted in $k_{j}$. Then, not to deviate from the event in (\ref{eq:thesis}), in between $k_{j}$ and $k_{j+1}$,  \emph{i.e.}, during a silent period of $\{u_i\}_{i=2}^N$, user $u_1$ must transmit, such that its index returns to zero.
 		
 		Denoting
 		\begin{align}
 		e(k_1^\ell) &\triangleq   \prod_{j=0}^{\ell-1}c(k_j,k_{j+1}),
 		\end{align}
 		and taking the average only over RV $K_\ell$, we get the following bound
 		\begin{align}
 		\expect{ e(k_1^{\ell-1},K_\ell)} & \ge\expect{1-(1-b^{-1-i_0})^{K_{\ell}-k_{\ell-1}-1} }  \nonumber\\
 		&\quad\cdot  \prod_{j=0}^{\ell-2}c(k_j,k_{j+1}) ,  \label{eq:back1}
 		\end{align}
 		where, recognizing that $K_{\ell}$ is a function of ${\bf X}_{k_{\ell-1}}$,  we can write
 		\begin{align}
 		&   \expect{1-(1-b^{-1-i_0})^{K_{\ell}-k_{\ell-1}-1}}  \nonumber\\
 		&= \expect{\expcnd{1-(1-b^{-1-i_0})^{K_{\ell}-k_{\ell-1}-1}}{{\bf X}_{k_{\ell-1}} }} . \label{eq:condX}
 		\end{align}
 		The RV $K_{\ell}-k_{\ell-1}$, when ${\bf X}_{k_{\ell-1}}=\x$, is Geometric-distributed with success probability $q(\x)$, and we have
 		\begin{align}
 		& \expcnd{1-(1-b^{-1-i_0})^{K_{\ell}-k_{\ell-1}-1}}{{\bf X}_{k_{\ell-1}}=\x } \nonumber \\
 		&=1-q(\x) \sum_{z=1}^\infty (1-q(\x ))^{z-1} (1-b^{-1-i_0})^{z-1} \nonumber\\
 		&\ge 1-q(\x) \sum_{z=0}^\infty  (1-b^{-1-i_0})^{z} \nonumber\\
 		&= 1-q(\x)\: b^{1+i_0} \label{eq:g_final}
 		\end{align}
 		where $q(\x)\ge 0$ is the probability that at least one user among $u_2^N$ transmits, which can be bounded as follows:
 		\begin{align}
 		q(\x) &= 1-\prod_{r=2}^N \left(1-b^{-x_{r}(k_{\ell-1})-i_0}\right)\nonumber\\
 		&\le  1- \left(1-b^{-x_{2}(k_{\ell-1})-i_0}\right)^{N-1}\nonumber\\
 		&\le (N-1)b^{-x_{2}(k_{\ell-1})-i_0}. \label{eq:bound_q}
 		\end{align}
 		Using~\eqref{eq:bound_q} into~\eqref{eq:g_final} yields
 		\begin{align}
 		& \expcnd{1-(1-b^{-1-i_0})^{K_{\ell}-k_{\ell-1}-1}}{{\bf X}_{k_{\ell-1}} } \nonumber \\
 		&\ge 1-(N-1)b^{-X_{2}(k_{\ell-1})+1} \nonumber \\
 		&\ge  1-(N-1)b^{-A_{\ell-1}+1}, \label{eq:g_finalss}
 		\end{align}
 		where we have used the lower bound to $x_2(k_\ell)$ in \eqref{eq: bounduno}:
 		\begin{align} A_\ell &= a_0 + \frac{\ell}{N-1}-(X_N(k_\ell)-X_2(k_\ell)) \nonumber\\
 		&\ge m + \frac{\ell}{N-1}-(X_N(k_\ell)-X_2(k_\ell)), \qquad \ell \ge 1,\end{align}
 		where the last inequality is due to $X_r(0)\ge X_2(0)=m$ for $r\ge 2$.
 		A bound to~\eqref{eq:condX} is:
 		\begin{align}
 		&   \expect{1-(1-b^{-1-i_0})^{K_{\ell}-k_{\ell-1}-1}}  \nonumber\\
 		&\ge 1-(N-1)b^{- m - \frac{\ell-1}{N-1}+1} \expect{b^{\Delta_N(k_{\ell-1})}} \nonumber\\
 		&\ge 1-(N-1)b^{- m - \frac{\ell-1}{N-1}+1} D
 		\end{align}
 		where the last inequality follows by result~\eqref{eq:distr12} of Lemma~\ref{Lemma:bound2}.
 		
 		Inequality \eqref{eq:back1}, then, becomes
 		\begin{align}
 		\expect{ e(k_1^{\ell-1},K_\ell)}& \ge \left( 1-(N-1)b^{- m - \frac{\ell-1}{N-1}+1} D\right) \nonumber\\
 		&\quad\cdot  \prod_{j=0}^{\ell-2}\left(1-(1-b^{-1-i_0})^{k_{j+1}-k_{j}} \right).  \label{eq:back12}
 		\end{align}
 		
 		Now we can take the expectation with respect to  $K_{\ell-1}$ and iteratively until $K_1$, and using the same bounding techniques as before we obtain
 		\begin{align}
 		\expect{ e(K_1^{\ell})}   &\ge \prod_{j=1}^{\ell}  \left( 1-(N-1)b^{- m - \frac{j-1}{N-1}+1} D\right) \label{eq:exp_e_kl}.
 		\end{align}
 		
 		Denoting $\gamma = (N-1) b^{-m+1} D$, for an infinitely long path we have
 		\begin{align}
 		\lim_{\ell\rightarrow\infty}\ln(\expect{ e(K_1^{\ell})}) &\ge \sum_{j=1}^\infty  \ln \left(1- \gamma\: b^{-\frac{j-1}{N-1}}\right)  \nonumber\\
 		&\stackrel{(a)}{\ge}-  \sum_{j=1}^\infty  \frac{ \gamma\: b^{-\frac{j-1}{N-1}} }{1-\gamma\: b^{-\frac{j-1}{N-1}}} \nonumber\\
 		&\stackrel{(b)}{\ge} - \frac{\gamma}{1-\gamma} \sum_{j=0}^\infty  b^{-\frac{j}{N-1}}  \nonumber\\
 		& > -\infty, \label{eq:limlog}
 		\end{align}
 		where step~$(a)$ follows by choosing $m$ such that $0<1-\gamma<1-\gamma\: b^{-\frac{j-1}{N-1}}<1$, and by the inequality $\log(1-x)\ge -x/(1-x)$ for $0\le x<1$. Step~$(b)$ follows by monotonicity of $b^{-\frac{j-1}{N-1}}$.
 	\end{IEEEproof}

 	% {corollary} --------------------------------------------------------
 	\begin{corollary} \label{cor:1} With the notation of Lemma \ref{Lemma1} we have
 		\begin{equation}\label{eq:thesis22}
 		\lim_{j\rightarrow \infty} \expect{\Pr\left(X_1(K_j)=0| \{X_1(K_v)=0,  {\cal C}(\X_{K_v})\}_{v=1}^{j-1} \right) }=1. \end{equation}
 	\end{corollary}
 	\begin{IEEEproof} Just note that~\eqref{eq:thesis22} is a necessary condition for the thesis~\eqref{eq:thesis} of Lemma~\ref{Lemma1}.
 	\end{IEEEproof}

 The results of this section prove that, if a sequence of $j$ consecutive collisions occur among users $\{u_i(0)\}_{i=2}^N$, the indexes of these users increase asymptotically with average rate $j/(N-1)$ and without spreading, \emph{i.e.}, process $\{X_N(k_j)-X_2(k_j)\}$ becomes asymptotically ergodic (Lemmas \ref{Lemma:bound2} and \ref{Lemma:bound3}). This has two consequences: first, after some collisions, further consecutive collisions occur with at most one of these users; second, since those transmissions become rare, after each collision index $X_1$  returns to zero well in advance with respect to a new transmission of $\{u_i(0)\}_{i=2}^N$ (Lemma \ref{Lemma:bound3}), a crucial point for stability issues.

\section{Main Results}\label{Sec:stability}

The conditions under which $\{\X_t \}$ is recurrent are given by the following theorem.

%-----------------------------------------------------------------------------------------------------------------------

% Now we can prove the following:
	\begin{theorem}\label{Th:recurrent}
		\MC{N} is recurrent for any $N\ge 2$, $b>1$, and $i_0>0$.
	\end{theorem}
	\begin{IEEEproof}
		First we need the following
		\begin{lemma} \label{lemma1}
			For any $N\ge 2$, $b>1$, and $i_0>0$,  the index  of a sample user $u$ returns to zero in a finite time.
		\end{lemma}		
		\begin{IEEEproof}
			We consider a set of ordered integers ${\bf k}=(k_1,k_2,\ldots,k_j)$, with $k_1\ge 0$ and $k_m < k_n$ for $m<n$, that represent the only time instants where user $u$ transmits. Let $a_u({\bf x}_t)=\prod_{i\ne u} (1-b^{-x_i(t)-i_0})$ be the probability  that no other user transmits at time $t$.
			
			Starting from any state $\x_{k_1}\in {\cal X}^{(N)}$, the probability that user $u$ successfully transmits at the $i$-th attempt, \emph{i.e.}, at time $K_i$, is
			\begin{align}
			P_i(\x_{k_1})&=\expcnd{a_u(\X_{K_i})}{\x_{k_1}} \nonumber\\
			&\ge (1-b^{-i_0})^{N-1},
			\end{align}
			therefore the expected return time to the state $x_u=0$ is bounded for any $N\ge 2$, $b>1$ and $i_0>0$.		
		\end{IEEEproof}
		As a consequence of the above lemma, values ${\bf x}_t$  are finite at any time for $b>1$ and $i_0>0$.
		
		Now we can prove the theorem by showing that the return time to state ${\bf x}=(0,0,\ldots,0)$ is finite. To the purpose, we denote by ${\cal Z}_i$ the subset of states with $i$ zero indexes and $N-i$ non-zero indexes.
		Clearly, $\{{\cal Z}_i\}_{i=0}^N$ is a partition of
		%${\cal S}^{(N)}$, that represents the \textit{macrostates} of
		${\cal X}^{(N)}$.  The number of such macrostates is finite, $N+1$, and in order to be back to ${\cal Z}_N$ within a finite time we must prove that $\{{\cal Z}_i\}_{i=0}^N$ is a closed communicating class, \emph{i.e.}, each ${\cal Z}_i$ is \emph{reachable} from any other ${\cal Z}_j$.

		The one-step transition probabilities between macrostates ${\cal Z}_i$ and ${\cal Z}_j$ depend on the departure state in the departure macrostate, \emph{i.e.},
		\begin{equation}
		p_{i,j}({\bf x}) \triangleq  \Pr({\bf X}_{t+1} \in {\cal Z}_{j}\: |\: {\bf X}_t = {\bf x}\in {\cal Z}_{i} ).
		\end{equation}

		The proof that $p_{i,i-k}({\bf x})>0, \ 1\le k \le i$, $\forall {\bf x}\in {\cal Z}_{i} $ is trivial, and is left to the reader. Here we prove that $p_{i,i+1}({\bf x})>0$,  $\forall {\bf x}\in {\cal Z}_{i} $  which suffices to prove that macrostates communicate. If ${\bf x}$ is the state at time $t$, we have
		\begin{align}
		p_{i,i+1}({\bf x}) &= \sum_{j=i+1}^{N}b^{-x_j-i_0} \prod_{k=1,\, k\ne j}^{N } (1-b^{-x_k-i_0}) \nonumber\\
		&=(1-b^{-i_0})^{i}\sum_{j=i+1}^{N}b^{-x_j-i_0}   \prod_{k=i+1,\, k\ne j}^{N} (1-b^{-x_k-i_0}). \label{eq: tran1}
		\end{align}
		In order to let ${\cal Z}_i$ communicate with  ${\cal Z}_{i+1}$ at any time, we must require that indexes $x_j$ can not increase without limit as time increases. We already observed after Lemma~\ref{lemma1} that values $x_j$ in (\ref{eq: tran1})  are finite at any time and, therefore,  probability (\ref{eq: tran1}) is always greater than zero for $b>1$ and $i_0>0$.
		
		\end{IEEEproof}

	%  theorem ---------------------------------------------------------
The following theorem proves the conditions for transience, and is based on the results remarked at the end of Sec.~\ref{Sec:preliminaries}. In particular, after some collisions, a further transmission of one among $\{u_i(0)\}_{i=2}^N$ finds $u_1$ with $X_1=0$, and can never have success if $i_0=0$.

\begin{theorem}\label{Th:transient2}
For $b>1$ and $i_0=0$ the MC \MC{N} is transient for any $N\ge 2$. Furthermore, the marginal distribution of $X_1$ asymptotically exists and is
\be \lim_{t \rightarrow \infty} \Pr(X_1(t)=0)=1. \label{diststaz1}\ee
\end{theorem}
\begin{IEEEproof}
Considering transmission times ${\bf k}=(k_1,k_2,\ldots,k_i,\ldots)$, defined as in Sec.~\ref{Sec:preliminaries}, and the set of events
\begin{align}
 {\cal P}({\bf k}) &= \{ {\cal C}(\x_{k_{1}}),{\cal C}(\x_{k_{2}}), \ldots,{\cal C}(\x_{k_i}), \ldots \},
\end{align}
where ${\cal C}(\x)$ denotes the collision event when the indexes are $\x$, we prove that $\Pr({\cal P}) > 0, $
\emph{i.e.}, the set of path events ${\cal P}$ where none of the users $\{u_i(0)\}_{i=2}^N$ ever has a success, has non-zero probability: Since with a positive probability they never experience success, their states never return to zero, hence the transience of the chain. To show this, write
\begin{align}
 &\Pr({\cal P}) =\expect{\Pr(\{{\cal C}(\X_{K_j})\}_{j=1}^\infty)} \nonumber\\
&\ge \expect{\Pr(\{{\cal C}(\X_{K_j}), X_1(K_j)=0\}_{j=1}^\infty)} \nonumber\\
%&= \expect{\prod_{j=1}^{\infty} \Pr({\cal C}(\X_{K_j}), X_1(K_j)=0|\{{\cal C}(\X_{K_l}), X_1(K_l)=0\}_{l=1}^{j-1})} \nonumber\\
%&= \expect{\prod_{j=1}^{\infty} \Pr({\cal C}(\X_{K_j})| X_1(K_j)=0) \Pr(X_1(K_j)=0|\{{\cal C}(\X_{K_l}), X_1(K_l)=0\}_{l=1}^{j-1})} \nonumber\\
&\stackrel{(a)}{=}\expect{\prod_{j=1}^{\infty}  \Pr(X_1(K_j)=0|\{{\cal C}(\X_{K_l}), X_1(K_l)=0\}_{l=1}^{j-1})} \nonumber\\
&>0
\end{align}
where~$(a)$ holds because $X_1(K_j)=0$ and $i_0=0$ cause a certain transmission of $u_1$ hence a collision at time instant $K_j$, and the last inequality is the result of Lemma~\ref{Lemma1}. Limit~\eqref{diststaz1} comes from Corollary \ref{cor:1}.
\end{IEEEproof}

\begin{theorem}\label{Th:necNx}
	For any $N\ge 2$, if  the average time to the first success among users $\{u_i(0)\}_{i=2}^N$ is finite, then $i_0>1/(N-1)$ and $b>1$.
\end{theorem}
\begin{IEEEproof}
Considering transmission times ${\bf k}$ as defined in Sec.~\ref{Sec:preliminaries}, and denoted by $S$ the index such that the first success among users $\{u_i(0)\}_{i=2}^N$ occurs at time $k_S$, the average time to the first success starting from state $\X_0 = \x$ can be written as
\begin{align}\label{eq:ts_suff_th6}
	t_s(\x) & \triangleq  \sum_{{\bf k},i}  k_{i}\: \Pr(S=i| {\bf K}={\bf k},\X_0=\x)\: \Pr({\bf K}={\bf k}|\X_0=\x) \nonumber \\
 & = \sum_{{\bf k}} \expcnd{k_S}{ {\bf k},\x} \: \Pr({\bf K}={\bf k}|\x)  .
	\end{align}
Following the lines of Theorem \ref{Th:transient2}, write
\begin{align}
& \Pr(S>i| {\bf k},\x) =\expect{\Pr(\{{\cal C}(\X_{k_j})\}_{j=1}^i)} \nonumber\\
&\ge \expect{\Pr(\{{\cal C}(\X_{k_j}), X_1(k_j)=0\}_{j=1}^i)} \nonumber\\
%&= \expect{\prod_{j=1}^{i} \Pr({\cal C}(\X_{k_j}), X_1(k_j)=0|\{{\cal C}(\X_{k_l}), X_1(k_l)=0\}_{l=1}^{j-1})} \nonumber\\
%&= \expect{\prod_{j=1}^{i} \Pr({\cal C}(\X_{k_j})| X_1(k_j)=0,\{{\cal C}(\X_{k_l}), X_1(k_l)=0\}_{l=1}^{j-1})  \Pr(X_1(k_j)=0|\{{\cal C}(\X_{k_l}), X_1(k_l)=0\}_{l=1}^{j-1})} \nonumber\\
&\stackrel{(a)}{\ge} \left( b^{-i_0}\right)^i\expect{\prod_{j=1}^{i}  \Pr(X_1(k_j)=0|\{{\cal C}(\X_{k_l}), X_1(k_l)=0\}_{l=1}^{j-1})} \nonumber\\
&\triangleq\left( b^{-i_0}\right)^i \beta_i, \label{eq:Sprime}
\end{align}
where step~$(a)$ follows by considering only the collision event caused by a transmission of user $u_1(k_j)$, that conditioned to $X_1(k_j)=0$ happens with probability $b^{-i_0}$.
	The RHS of~\eqref{eq:Sprime} represents the distribution $\Pr(S'>i)$ of a RV $S'$ that is stochastically dominated by $S$, \emph{i.e.}, $\Pr(S>i)\ge \Pr(S'>i)$ for all $i$.
Therefore, averaging any non-decreasing functions $g(\cdot)$ with respect to $S'$ provides a lower bound to averaging with respect to $S$ \cite{quirk1962}, and we can write:
\begin{align}%\label{eq:ts_suffa}
	 &\Exp[k_S| {\bf k},\x] =  \sum_{i=1}^{\infty}  k_{i}\: \Pr(S=i| {\bf k},\x) \nonumber \\
 &\ge  \sum_{i=1}^{\infty}  k_{i}\: \Pr(S'=i)
  =  \sum_{i=1}^{\infty}  k_{i}\: \left( \left( b^{-i_0}\right)^{i-1} \beta_{i-1}-\left( b^{-i_0}\right)^i \beta_i \right) \nonumber
	\end{align}
where we have used
\begin{align} \Pr(S'=i)&=\Pr(S'>i-1)-\Pr(S'>i)\nonumber\\ &=\left( b^{-i_0}\right)^{i-1} \beta_{i-1}-\left( b^{-i_0}\right)^i \beta_i. \label{eq:pdfS}
\end{align}
Using~\eqref{eq:pdfS} into~\eqref{eq:ts_suff_th6} yields
\begin{align}\label{eq:ts_suffca}
	t_s(\x) \ge & \sum_{i=1}^{\infty}\Pr(S'=i) \sum_{{\bf k}}  k_{i}\: \Pr({\bf K}={\bf k}|\x) \nonumber \\
= & \sum_{i=1}^{\infty}\left( \left( b^{-i_0}\right)^{i-1} \beta_{i-1}-\left( b^{-i_0}\right)^i \beta_i \right) \expcnd{K_i}{\x}.
	\end{align}
Conditioned to $\X_{K_{j-1}}$, RV $K_j-K_{j-1}$ is Geometric-distributed with success probability
\begin{align} q(\X_{K_{j-1}})&=1-\prod_{r=2}^{N}(1-b^{-X_{r}(K_{j-1})-i_0}) \nonumber\\
&\le 1-(1-b^{-X_{2}(K_{j-1})-i_0})^{N-1} \nonumber\\
&\le (N-1) b^{-X_{2}(K_{j-1})-i_0}\nonumber\\
&\le (N-1) b^{-x-\frac{j-1}{N-1}+\Delta_N(K_{j-1})-i_0} \end{align}
where the last step follows by assuming that $X_2(0)=x$ and by Lemma~\ref{Lemma:bound1}.
Hence we have:
\begin{align}
&\expcnd{K_i}{ \x } = \sum_{j=1}^{i}\expcnd{ K_j - K_{j-1}}{\x} \nonumber\\
%&= \sum_{j=1}^{i}\expcnd{ \expcnd{K_j - K_{j-1}}{\X_{K_{j-1}}}  }{\x} \nonumber\\
&=\sum_{j=1}^{i} \expcnd{\left(q(\X_{K_{j-1}})\right)^{-1}}{\x} \label{eq:Ki_q} \\
 &\ge \frac{1}{N-1}\sum_{j=1}^{i} b^{x+\frac{j-1}{N-1}+i_0}\expcnd{ b^{-\Delta_N(K_{j-1})}}{\x}  \nonumber \\
&\stackrel{(a)}{\ge} \frac{1}{N-1}\sum_{j=1}^{i} b^{x+\frac{j-1}{N-1}+i_0}\left(\expcnd{ b^{\Delta_N(K_{j-1})}}{\x}\right)^{-1} \nonumber\\
&\stackrel{(b)}{\ge}\frac{ b^{x+i_0-D}}{N-1}\frac{ b^{i/(N-1)}-1}{b^{1/(N-1)}-1 }\label{eq:ineq02c}
\end{align}
where $(a)$ holds by Jensen's inequality, and $(b)$
 by Lemma~\ref{Lemma:bound2} where $\expcnd{b^{\Delta_N(K_{j-1})}}{\x}\le D<\infty$.
%\begin{align}
%\expcnd{K_i}{\x}
%&\ge \frac{ b^{x+i_0-D}}{N-1}\sum_{j=1}^{i}b^{\frac{j-1}{N-1}}\nonumber \\
%&= \frac{ b^{x+i_0-D}}{N-1}\frac{ b^{i/(N-1)}-1}{b^{1/(N-1)}-1 }.\label{eq:ineq02c}
%\end{align}
Finally, bound \eqref{eq:ineq02c} can be used in \eqref{eq:ts_suffca} to give
\begin{align} %\label{eq:ts2}
&t_s({\bf x}) \ge  \frac{ b^{x+i_0-D}}{N-1}   \sum_{i=1}^{\infty}\left( b^{i_0} \beta_{i-1}- \beta_i \right)\left( b^{-i_0}\right)^i\frac{ b^{i/(N-1)}-1}{b^{1/(N-1)}-1 }. \nonumber
\end{align}
Asymptotically, probability $\beta_i$ becomes one, as from Corollary~\ref{cor:1},  and the convergence of the above summation is assured only if $b>1$ and $i_0>1/(N-1)$.
\end{IEEEproof}

%% theorem------------------------------------------------------------

\begin{theorem}\label{Th:suffbis}
	If $b>1$ and $i_0>1/(N-1)$, then  the average time to the first success among users $\{u_i(0)\}_{i=2}^N$ is finite for any $N\ge 2$.
\end{theorem}
\begin{IEEEproof}
	We follow the same lines of Th.~\ref{Th:necNx}, this time evaluating an upper bound to $t_s(\x)$.
%\begin{align}\label{eq:ts_suff_th7}
%	t_s(\x) & = \sum_{{\bf k}} \expcnd{k_S}{ {\bf k}, \x} \: \Pr({\bf K}={\bf k}|\x)  .
%	\end{align}
Denoting ${\cal C}'(\X_{k_j})$ the  collision event caused by a transmission of $u_1(k_j)$, and ${\cal C}''(\X_{k_j})$ the  collision event caused by a transmission of other users, we can write
 	\begin{align}
& \Pr(S>i| {\bf k}, \x) =\expect{\Pr(\{{\cal C}(\X_{k_j})\}_{j=1}^i)} \nonumber\\
%&= \expect{\prod_{j=1}^{i} \Pr({\cal C}(\X_{k_j})|\{{\cal C}(\X_{k_l})\}_{l=1}^{j-1})} \nonumber\\
&= \expect{\prod_{j=1}^{i} \Pr\left({\cal C}''(\X_{k_j}) \cup \left(\overline{{\cal C}''(\X_{k_j})} \cap {\cal C}'(\X_{k_j})\right) |\{{\cal C}(\X_{k_l})\}_{l=1}^{j-1}\right)}  \nonumber\\
&\le \expect{\prod_{j=1}^{i} \left(\Pr\left({\cal C}''(\X_{k_j}) |\{{\cal C}(\X_{k_l})\}_{l=1}^{j-1}\right)+b^{-i_0}\right)}   \label{eq:csec}
\end{align}
where the inequality follows by union bound and by $\Pr({\cal C}'(\X_{k_j}))\le b^{-i_0}$.
For any $\varepsilon>0$, we can always choose an initial state $\x$ such that
\be \Pr\left({\cal C}''(\X_{k_j}) |\{{\cal C}(\X_{k_l})\}_{l=1}^{j-1}\right) < \varepsilon, \qquad \forall j, \ee
and \eqref{eq:csec} becomes
 	\begin{align}
& \Pr(S>i| {\bf k},\x) \le  \left(\varepsilon+b^{-i_0}\right)^i.\label{eq:pI_K_suffa}
	\end{align}
The RHS of~\eqref{eq:pI_K_suffa} represents the distribution $\Pr(S''>i )$ of a RV $S''$ that stochastically dominates $S$.
Therefore, averaging any non-decreasing function  $g(\cdot)$ with respect to $S''$ provides an upper bound to averaging with respect to $S$:
\begin{align}\label{eq:ts_suffa_th7}
	&\expcnd{k_S}{{\bf k},\x}  = \sum_{i=1}^{\infty}  k_{i}\: \Pr(S=i| {\bf k},\x) \nonumber \\
 &\le \sum_{i=1}^{\infty}  k_{i}\: \Pr(S''=i) = \sum_{i=1}^{\infty}  k_{i}\:  (1-\varepsilon-b^{-i_0})(\varepsilon+b^{-i_0})^{i-1},
	\end{align}
where we have used
\begin{align}
	\Pr(S''=i) &= \Pr(S''>i-1)-\Pr(S''>i) \nonumber \\
& =(1-\varepsilon-b^{-i_0})(\varepsilon+b^{-i_0})^{i-1}. \label{eq:pI_K_suff}
	\end{align}	

Using~\eqref{eq:ts_suffa_th7} we get the bound
\begin{align}\label{eq:ts_suffcax}
	t_s(\x) \le
	%& \sum_{i=1}^{\infty} (1-\varepsilon-b^{-i_0})(\varepsilon+b^{-i_0})^{i-1}\sum_{{\bf k}}  k_{i}\: \Pr({\bf K}={\bf k}|\x) \nonumber \\
 (1-\varepsilon-b^{-i_0})\sum_{i=1}^{\infty}(\varepsilon+b^{-i_0})^{i-1} \expcnd{K_i}{\x}  .
	\end{align}
To show that the above series converges we must prove that
\be \lim_{i \rightarrow \infty}\expcnd{K_i}{\x}^{1/i}(\varepsilon+b^{-i_0})<1. \label{eq:ts_suffxxx} \ee

Conditioned to $\X_{K_{j-1}}$, RV $K_j-K_{j-1}$ is Geometric-distributed with success probability
\begin{align}
&q(\X_{K_{j-1}})  = 1-\prod_{r=2}^N (1-b^{-X_{r}(K_{j-1})-i_0})\nonumber \\
&\quad\ge b^{-X_{2}(K_{j-1})-i_0} \ge b^{-a_0 - \frac{j-1}{N-1} - \Delta_C(K_{j-1})-i_0} \label{eq:bound_q}
\end{align}
where the last step follows by Lemma~\ref{Lemma:bound1}.
Thanks to~\eqref{eq:Ki_q} and~\eqref{eq:bound_q}, the average time up to $K_i$ can be upper-bounded as
\begin{align}\label{eq:ts2y}
\expcnd{K_i}{\x}
%&= \sum_{j=1}^{i}\expcnd{q(\X_{K_{j-1}})^{-1}}{\x} \nonumber \\
&\le  \sum_{j=1}^{i} b^{a_0 + \frac{j-1}{N-1} +i_0}\expcnd{b^{ \Delta_C(K_{j-1})}}{\x} = \zeta\: b^{i/(N-1)} \nonumber
\end{align}
where $\zeta$ is independent of $i$ and we have used the fact that, by  Lemma \ref{Lemma:bound3}, $\expect{b^{\Delta_C(K_{j-1})}}$ is bounded by a finite constant.
Then, \eqref{eq:ts_suffxxx} is verified if $b>1$ and
\be (\varepsilon+ b^{-i_0}) b^{1/(N-1)} <1, \ee
for any $\varepsilon>0$, that is,  if $i_0 > \frac{1}{N-1}$.
 \end{IEEEproof}	

We now are in position to give the conditions for ergodicity. If the chain is not ergodic the joint distribution becomes asymptotically identically zero.  However, there are some conditions on $i_0$ that allow some marginal asymptotic distributions to exist even if the joint distribution does not exist, as we have seen for $X_1$ in the transient case (Th.~\ref{Th:transient2}). Therefore, we focus on the behavior of marginal process $\{X_r(t)\}$, which is clearly not Markovian; nevertheless, with a small abuse of language, we say that $\{X_r(t)\}$ is recurrent if its first return to zero occurs in a finite time, and that is positive recurrent if this time is finite in average.

We saw, in Th.~\ref{Th:transient2}, that $\{X_1(t)\}$ is positive recurrent for $b>1$ and $i_0\ge 0$. For the other components we have:

\begin{theorem}\label{Th:marginal}
\MC{N} is ergodic if and only if $b>1$ and $i_0>1$. Furthermore, $\{X_r(t)\}$, $r>1$, is positive recurrent if and only if $b>1$ and $i_0>1/(N-r+1)$, and all the marginal distributions up to $r$-th component exist for the same condition.
\end{theorem}
\begin{IEEEproof}
Starting at time zero, let $W_r^{(N)}$, $1 \le r \le N$ be the first time instant at which $X_r$ returns to zero in the chain with $N$ users. By definition, it is clear that $X_r$ can reach zero only if $X_s=0, \ s=1,\ldots, r-1$ and, therefore, we have  $\Exp[W_{r}^{(N)}]\ge \Exp[W_{r-1}^{(N)}]$. For $r=2$ the thesis is proved by Theorems  \ref{Th:necNx} and \ref{Th:suffbis}, which assure that $W_2^{(N)}$ has a finite average.

Let now $W_2'^{(N)}$ be the time instant where  $X_2$ returns to zero, and such that from there $X_1$ remains at zero until time $W_3^{(N)}$, where $X_3$ reaches zero too - notice that all sample paths must contain this event if the chain is recurrent, \emph{i.e.}, reaches the all zero state with probability one. Since $X_1$ is not allowed to leave state zero between $W_2'^{(N)}$ and $W_3^{(N)}$, it must not transmit at times $K_j$ contained in this time interval, \emph{i.e.}, $u_1(W_2'^{(N)})$ can not influence the behavior of the process $(X_2,X_3,\ldots,X_N)$ until $W_3^{(N)}$. This means that between $W_2'^{(N)}$ and $W_3^{(N)}$ the system behaves as a system with $N-1$ users where $u_1(W_2'^{(N)})$ is discarded, meaning that $W_3^{(N)}-W_2'^{(N)}$ in the chain with $N$ users presents the same average properties as $W_2^{(N-1)}$ in the chain with $N-1$ users. The time $W_2^{(N-1)}$, by Theorems  \ref{Th:necNx} and \ref{Th:suffbis}, has a finite average for $i_0>1/(N-2)$; therefore, also  $W_2'^{(N)}+(W_3^{(N)}-W_2'^{(N)})=W_3^{(N)}$ has finite average.

 We can repeat the argument for $W_r^{(N)},\, 3< r \le N$, proving the first part of the thesis, since the condition for ergodicity is clearly  $ \Exp[W_{N}^{(N)}]<\infty$, \emph{i.e.}, $i_0>1$.

Let now assume that  $ \Exp[W_{r}^{(N)}]<\infty, \ r<N$, and  $ \Exp[W_{N}^{(N)}]=\infty$. This means that  within period $W_N^{(N)}$, $X_r$,   $r<N$, returns to zero infinitely many times, so that one period $W_N^{(N)}$ can be seen as  composed by infinitely many periods, indexed by $k$, whose $k-$th length is $W_r^{(N)}(k)$. Since the fraction of time where $X_r=0$ in each period $W_r^{(N)}(k)$ is at least one for any $k$,  the average, over the entire time axis, of the time spent in state $X_r=0$ is greater than zero, and this average tends to $\Pr(X_r=0)>0$, showing that the distribution of $X_r$ exists.
\end{IEEEproof}	

By Th.~\ref{Th:marginal}, for $b>1$ and $i_0>1/(N-1)$, at least $X_2$ is positive recurrent, its transmissions and collisions occur infinitely many times so that the throughput of the system can not be one. However, for $ 0 <i_0 \le 1/(N-1)$, all components $X_r, \ r>1$ are null recurrent, meaning that $X_1$ returns to zero infinitely many times before $X_2$ does the same. This means we are in the conditions where \eqref{eq:thesis22}  holds and the throughput is one. This result, together with the result of Corollary \ref{cor:1} finally yields:

\begin{theorem}\label{Th:throughput}
	If $b>1$ and $i_0 \le \frac{1}{N-1}$, the throughput of \MC{N} is one.
\end{theorem}

Theorems \ref{Th:marginal} and \ref{Th:throughput} shows that \MC{N}, for $b>1$ and $i_0<1$,  can operate in situations where some users are locked out, while the positive throughput is shared among the remaining users.

	\section{General Stability}\label{Sec:genstability}

Let denote by ${\cal S}$ a system of $N$ queues where packets arrive  according to a Poisson process of rate $\lambda$, equally subdivided among queues,  and served by S-Aloha with EB. Let $\lambda_0$ be the throughput with a positive recurrent EB and queues in saturation. This throughput can be numerically evaluated as shown in \cite{infocom-2016}.  Then we have:
\begin{theorem}\label{Th:queue} The system ${\cal S}$, with a positive recurrent EB with throughput $\lambda_0$,  is stable if and only if $\lambda < \lambda_0$. \end{theorem} 

\begin{IEEEproof} Let consider  a new system ${\cal S}'$, equal to ${\cal S}$ but with queues that never empty,  \emph{i.e.}, a user that successfully transmits and is alone in the system, instead of leaving repeats its transmission and leaves only when, in doing this, the queue is not empty. Therefore, EB works in saturation condition. Let now focus on queue~$1$: Arrivals at this queue occur with rate $\lambda/N$ while the \emph{service time}  is clearly independent of all queues content, and has rate $\lambda_0/N$. Each single queue behaves as the \emph{modified}  $M/G/1$ system in Appendix~\ref{sec:appendix_A} where, as above, the queue is never left empty. Lemma~\ref{mg1} in Appendix~\ref{sec:appendix_A} shows that the modified $M/G/1$ system, and thus ${\cal S}'$,  is stable if and only if $\lambda <\lambda_0$. Clearly, the queue of system ${\cal S}$ can not be stable if $\lambda \ge \lambda_0$. Assume now that the queues are unstable for some  $\lambda< \lambda_0$; then  queues  build up and never empty again. But in this case the system behaves exactly as ${\cal S}'$, which  is stable for $\lambda< \lambda_0$, a contradiction. Therefore, the queue of system ${\cal S}$ is stable for $\lambda< \lambda_0$. \end{IEEEproof}

It is clear by Th.~\ref{Th:queue} that  ${\cal S}$ and ${\cal S}'$ have the same limiting throughput $\lambda_0$.

The decoupling between EB and queues is not possible if EB in saturation is not stable. This does not imply that queues can not be stable and that a positive throughput can not exist even in this case. As an example, let refer to $N=2$, in which case unstable EB means that in saturation conditions one of the two stations, say station A, captures the channel and keeps on transmitting as long as the queue is not empty.  During this capture period, the other queue, queue B, can not be served and its content builds up. When  queue A empties capture ceases, a long period occurs before station B transmits again, owing to its high index, after which a period of mixed transmissions occurs up to when station B captures the channel in its turn, and the mechanism repeats. We see, therefore, that on the channel, periods of capture alternate between the two stations with a switching period in between. If the switching periods were bounded, compared to the capture periods, then the maximum throughput would be $100\%$, as the arrival rate can be pushed up to make the queue length as great as wanted. In practice, the switching-period length is not bounded, because the  index value reached by the cut-out station is not bounded. This means that the maximum throughput can not be $100\%$. As an example, in \cite{backoff-1} we have argued that the maximum throughput in the $N=2$ case, under Bernoulli arrivals and $i_0=0$, is approximately 0.6096. However, the general solution in this case is still an open problem.

%-------------------------------------------------------
\section{Conclusions}\label{Sec:conclusions}

In this paper we have provided the stability/instability conditions for slotted Aloha under the backoff model with exponential law $i\mapsto b^{-i-i_0}$, when transmission buffers are always full. When referring to a system with queues and Poisson arrivals, we have proven that the system is stable whenever the exponential backoff (EB) in saturation is stable with throughput $\lambda_0$ and the input rate $\lambda$ is upper-bounded as $\lambda<\lambda_0$. Future work directions are to find general stability conditions, \emph{i.e.}, maximum throughput, when EB in saturation is not stable, and to investigate more general backoff laws.

\appendices

\section{}\label{App:MC}

Let consider the homogeneous Markov Chain $(\Delta_N(n),\widetilde{\X}_n)=(\Delta_N(n),X_2(n),\ldots,X_{N-1}(n))$, where $\Delta_N(n)$ can at most increase or decrease by one at each step, with transition probabilities $p(\Delta_N (n+1)=d, \x_{n+1}|\Delta_N (n)=\delta,\x_n)$, and let assume that we change the transition probabilities into
$p'(\Delta_N (n+1)=d, \x_{n+1}|\Delta_N (n)=\delta,\x_n)$, giving rise to the Markov Chain $(\Delta'_N(n),\X'_n)=(\Delta'_N(n),X'_2(n),\ldots,X'_{N-1}(n))$, always having $|\Delta'_N(n+1)-\Delta'_N(n)| \le 1$.
Denoting
\begin{align}  p(d|\delta,\x_n)& =\sum_{\x_{n+1}} p(\Delta_N (n+1)=d, \x_{n+1}|\Delta_N (n)=\delta,\x_n) \\
p'(d|\delta,\x_n)& =\sum_{\x_{n+1}} p'(\Delta_N (n+1)=d, \x_{n+1}|\Delta_N (n)=\delta,\x_n)
\end{align}
for all $d,\delta,\x_n$, and
\begin{align}  p(d|\delta) &= \sum_{\x_{n}}  p(d|\delta,\x_n)\pi(\x_n|\delta), \qquad\forall d,\delta, \\
p'(d|\delta) &= \sum_{\x_{n}}  p'(d|\delta,\x_n)\pi'(\x_n|\delta), \qquad\forall d,\delta,\end{align}
where $\pi(\x_n|\delta)$ and $\pi'(\x_n|\delta)$ are the  conditional distributions of $\x_n$ in the two processes respectively. We  have

\begin{lemma}\label{lem:delta_dominanza} If the change in the transitions probabilities is such that
	\begin{align}  p'(d|\delta,\x_n)& \ge p(d|\delta,\x_n), \qquad d >\delta, \label{eq: delta1} \\
	p'(d|\delta,\x_n)& \le p(d|\delta,\x_n), \qquad d <\delta, \label{eq: delta2} \\
	p'(d|\delta) & \le 1/2, \qquad \forall \delta\ge 0, \label{eq: delta3}
	\end{align}
	for all $\x_n$, then the marginal process $\{\Delta'_N(n)\}$ stochastically dominates $\{\Delta_N(n)\}$ for all $n$, \emph{i.e.}, starting from the same initial condition we have
	\be \Pr(\Delta'_N(n)>\delta) \ge \Pr(\Delta_N(n)>\delta), \qquad \forall \delta,n \ge 0. \ee
\end{lemma}
\begin{IEEEproof}
	The assumptions imply the following inequalities
	\begin{align}  p'(d|\delta)& \ge p(d|\delta), \qquad d >\delta, \nonumber \\
	p'(d|\delta)& \le p(d|\delta), \qquad d <\delta. \nonumber
	\end{align}
	Since the above are distributions, in particular their sum over $d$ is one,  we can write
	\begin{align} &  p'(d|\delta)= p(d|\delta)+v(d|\delta) , \qquad\forall d,\delta, \nonumber \end{align}
	\be \sum_{d=0}^\infty v(d|\delta)=0, \qquad \forall \delta, \label{eq:cos1} \ee
	with $v(d|\delta) \ge 0$ for $d \ge \delta$, and $v(d|\delta)<0$ for $d < \delta$. Then, because of the properties above, we can write \begin{align}   F'_{n+1}(d|\delta) & = \sum_{y=d+1}^\infty p'(d|\delta)
	= \sum_{y=d+1}^\infty  \big(p(d|\delta)+  v(d|\delta)\big) \nonumber \\
	& =  F_{n+1}(d|\delta)+\sum_{y=d+1}^\infty  v(d|\delta)  \nonumber \\ & \ge F_{n+1}(d|\delta), \qquad \forall d,\delta,  \label{eq: domin1} \end{align}
	where the last inequality comes from the constraint \eqref{eq:cos1} over $v(\cdot|\delta)$. \eqref{eq: domin1} shows that $\Delta_N'(n)$  stochastically dominates  $\Delta_N(n)$, and we can use the property of stochastic dominance by which, if $g(\delta)$ is a weakly increasing function of $\delta$, the following holds \cite{quirk1962}
	\be \sum_{\delta}   g(\delta)  \Pr(\Delta_N'=\delta)   \ge \sum_{\delta}   g(\delta)  \Pr(\Delta_N=\delta) \label{eq:prop} . \ee
	Denoting
	\be \Delta F_{n+1}(d|\delta) \triangleq  F'_{n+1}(d|\delta) - F_{n+1}(d|\delta)\ge 0, \qquad \forall d \ge 0, \label{eq:prop2} \ee
	we can write
	\begin{align}  F_{n+1}(d)& =\sum_{\delta}F_{n+1}(d|\delta)\Pr(\Delta_N(n)=\delta) \\
	& =\sum_{\delta}F'_{n+1}(d|\delta)\Pr(\Delta_N(n)=\delta)  \nonumber\\
	&\quad-\sum_{\delta}\Delta F_{n+1}(d|\delta)\Pr(\Delta_N(n)=\delta) \nonumber \\
	&\stackrel{(a)}{\le} \sum_{\delta}F'_{n+1}(d|\delta)\Pr(\Delta'_N(n)=\delta) \nonumber\\
	&\quad -\sum_{\delta}\Delta F_{n+1}(d|\delta)\Pr(\Delta'_N(n)=\delta)   \nonumber  \\
	&=F'_{n+1}(d) -\sum_{\delta}\Delta F_{n+1}(d|\delta)\Pr(\Delta_N(n)=\delta)  \nonumber \\
	&\stackrel{(b)}{\le} F'_{n+1}(d)  \nonumber
	\end{align}
	where inequality $(a)$ comes from \eqref{eq:prop}, and the fact that $F'_{n+1}(d|\delta)$ is an increasing function of $\delta$. In fact, because the increments of $\Delta$ and $\Delta'$ takes place one unit at a time, we have
	\begin{align}
	F'_{n+1}(d|\delta)=\left\{\begin{array}{cc}
	0 & \delta < d,\\
	p'(\delta+1|\delta) & \delta=d,\\
	1-p'(\delta+1|\delta) & \delta = d+1,\\
	1 & \delta>d+1,
	\end{array}\right.
	\end{align}
	%
	%\begin{align}   F'_{n+1}(d|\delta)&= 0, \qquad   & \delta < d, \\
	%F'_{n+1}(d|\delta)&= p'(\delta+1|\delta), \qquad  & \delta = d, \label{eq: d}\\    F'_{n+1}(d|\delta)&= 1-p'(\delta+1|\delta), \qquad  & \delta = d+1, \label{eq: d}\\
	%                F'_{n+1}(d|\delta)&= 1, \qquad   &\delta >d+1,
	%\end{align}
	where the increasing property comes from \eqref{eq: delta3}.
	The above result proves that  if $\Delta_N'(n)$  stochastically dominates $\Delta_N(n)$, then also $\Delta_N'(n+1)$ stochastically dominates $\Delta_N(n+1)$. Then, starting from the same initial condition, the lemma is proven.
\end{IEEEproof}

 \section{}\label{sec:appendix_A}

Let consider an $M/G/1$ queue with Poisson arrivals at rate $\lambda$ and average service time $1/\lambda_0$. Let also consider a \emph{modified}  $M/G/1$ system where the queue never empties, \emph{i.e.}, a user that successfully transmits and is alone in the system, instead of leaving repeats its transmission and leaves only when, in doing this, the queue is not empty.

\begin{lemma}\label{mg1} The modified $M/G/1$ system is stable if and only if the $M/G/1$ system is stable, \emph{i.e.}, $\lambda <\lambda_0$. \end{lemma} \begin{IEEEproof} The proof is based on considering  the distribution of RV $R_k$, defined as the system content at time instant $t_k$ when packet $k$ leaves the system (this approach is followed by many books; see for example \cite{cooper}). $\{R_k\}_{k\ge 1}$ is an MC whose asymptotic distribution is proved to be equal to the distribution seen upon arrivals, which, in turn, is equal to the asymptotic distribution in time, owing to the  ``Poisson arrivals see time averages'', or PASTA, property~\cite{wolff}. This distribution, in $M/G/1$ systems, exists  if and only if $\lambda <\lambda_0$, and its transition matrix can be written as 
	 \begin{equation} P = [p_{ij}] = \left[ \begin{array}{ccccc} \zeta_{0} & \zeta_{1} & \zeta_{2} & \zeta_{3} & \ldots \\ \zeta_{0} & \zeta_{1} & \zeta_{2} & \zeta_{3} & \ldots \\ 0     & \zeta_{0} & \zeta_{1} & \zeta_{2} & \ldots \\ 0     & 0     & \zeta_{0} & \zeta_{1} & \ldots \\
	 \vdots & & & \ddots & \ddots \\  \end{array} \right] \label{eq: matrmg}, \end{equation} $i,j \ge 0$, where $\zeta_j$ represents the probability of $j$ arrivals between two consecutive departures from the system. It is easily recognized that the modified system presents the transition matrix \begin{equation} P' = [p_{ij}'] = \left[ \begin{array}{ccccc} \zeta_0+\zeta_{1} & \zeta_{2} & \zeta_{3} & \zeta_4 &\ldots \\ \zeta_{0} & \zeta_{1} & \zeta_{2} & \zeta_3 & \ldots \\ 0     & \zeta_{0} & \zeta_{1} & \zeta_2 & \ldots \\ 0  &0   & \zeta_{0} & \zeta_{1} & \ldots \\
	 \vdots & & & \ddots & \ddots \\ \end{array} \right], \label{eq: matrmg} \end{equation}
	 $i,j\ge 1$. 
	To prove stability and instability conditions we refer to  Foster's Theorem \cite{foster} and  Kaplan's Theorem \cite{kaplan} respectively. Both theorems are based on some conditions about the average drift of a potential function $f$: \be \expcnd{f(R_{k+1})-f(R_k)}{R_k=r} \ee for $r>r_0$, for some $r_0\ge 0$. We see that the average drift  for $r_0>0$ is equal for both systems. Hence, conditions on the existence of the distribution are the same in both systems. \end{IEEEproof}

\bibliographystyle{IEEEtran}
\bibliography{biblio1}

\end{document}